\documentclass[reprint, superscriptaddress, amsmath, amssymb, aps, prc, nofootinbib]{revtex4-2}
\usepackage{graphicx}
\usepackage{mathtools}
\usepackage[hidelinks]{hyperref}
\usepackage{multirow}

\begin{document}

\title{Wigner distribution modeling of nuclear state in NuWro Monte Carlo event generator}

\author{Rwik Dharmapal Banerjee}\email{rwikdharmapal.banerjee@phd.uwr.edu.pl}
\affiliation{
Institute of Theoretical Physics, University of Wroc{\l}aw,
plac Maxa Borna 9, 50-204 Wroc{\l}aw, Poland
}

\author{Alexis Nikolakopoulos}\email{anikolak@uw.edu}
\affiliation{
Department of Physics, University of Washington,
Seattle, Washington 98195-1560, USA
}

\author{Jan T. Sobczyk}
\affiliation{
Institute of Theoretical Physics, University of Wroc{\l}aw,
plac Maxa Borna 9, 50-204 Wroc{\l}aw, Poland
}

\author{Hemant Prasad}
\affiliation{
Institute of Theoretical Physics, University of Wroc{\l}aw,
plac Maxa Borna 9, 50-204 Wroc{\l}aw, Poland
}

\author{Krzysztof M. Graczyk}
\affiliation{
Institute of Theoretical Physics, University of Wroc{\l}aw,
plac Maxa Borna 9, 50-204 Wroc{\l}aw, Poland
}

\author{Jos\'e L. Bonilla}
\affiliation{
Institute of Theoretical Physics, University of Wroc{\l}aw,
plac Maxa Borna 9, 50-204 Wroc{\l}aw, Poland
}

\author{Beata E. Kowal}
\affiliation{
Institute of Theoretical Physics, University of Wroc{\l}aw,
plac Maxa Borna 9, 50-204 Wroc{\l}aw, Poland
}

\date{\today}

\begin{abstract}
We develop a method to use the Wigner distribution (WD) description of the nuclear
initial state of $^{12}$C, $^{16}$O, $^{40}$Ar, and $^{56}$Fe for quasielastic scattering
in the \textsc{NuWro} Monte Carlo event generator. We propose two different algorithms to deal with the problem of negative values of the WDs. In our treatment, finite-nucleus WDs provide a
correlation between the position and the momentum of bound nucleons that is very
different from what is assumed in the standard approaches based on the Fermi gas model
and spectral functions. We show that the use of WDs leads to effects that can be
measured experimentally and may be relevant for future neutrino oscillation experiments.  
\end{abstract}

\maketitle

\section{Introduction}

With the advent of the new generation of long-baseline neutrino oscillation experiments: HyperKamiokande~\cite{Hyper-Kamiokande:2018ofw} and DUNE~\cite{DUNE:2020lwj}, it is becoming more important to construct reliable models of neutrino-nucleus scattering. The pattern of neutrino oscillations depends on neutrino energy, which is not known exactly, and must be reconstructed using information from particles resulting from interactions~\cite{CLAS:2021neh}. Theoretical models must be implemented in Monte Carlo (MC) event generators, and one of the biggest challenges is that in their predictions they must cover the full available phase space. 

Neutrino event generators model the complex hadronic final-states produced in neutrino interactions by making use of a two-step approach. A primary interaction on a bound nucleon which produces some hadrons is followed by the modeling of final state interactions (FSI) of said hadrons by making use of an intranuclear cascade (INC) model. 

The INC picture goes
back to Serber~\cite{Serber:1947zza} and to the MC cascade calculations of Metropolis \textit{et al.}~\cite{Metropolis:1958sb,Metropolis:1958wvd} and Bertini~\cite{Bertini:1963zzc}. Today it is the standard tool for hadron rescattering in neutrino event generators, e.g.\ \textsc{NuWro}~\cite{Golan:2012wx,
Niewczas:2019fro}, \textsc{GENIE}~\cite{Andreopoulos:2009rq} and
\textsc{NEUT}~\cite{Hayato:2021heg}. Within this model, an important degree of freedom is where the hadrons are initially placed in the INC. The simplest approach is to select this position based on the total nuclear density (or alternatively by considering the position-momentum correlation from the local-density approximation (LDA), i.e. the local Fermi gas (LFG) model). 

However, there are certain nuclear effects that can only be included by a more precise prescription of where the hadrons originate inside the nucleus.
 For example, calculations for single-nucleon knockout~\cite{Lava04, Ryckebusch05, Nikolakopoulos24, Butkevich25, RGJ:2026Transparency} show that the strength of final-state interactions (measured by nuclear transparency) depends strongly on the specific residual system that is produced. As shown recently in Ref.~\cite{RGJ:2026Transparency}, this effect can be naturally replicated by introducing nucleons in the INC according to the radial distribution function associated with the hole-state. While this might not readily be observable in neutrino experiments since all final states are summed over~\cite{Nikolakopoulos:2022qkq}, including this effect in event generators is crucial for comparison of the two-step approximation to electron-induced nucleon knockout reactions. 
 
 Furthermore, the JSNS$^2$ measurement~\cite{JSNS2:2024bzf} of visible-energy in mono-energetic neutrino interactions allows one to partly disentangle the contribution of different single-particle states in ${}^{12}$C. As we will show, the modification of the starting position of the nucleon in the \textsc{NuWro} INC leads to about 10\% effects for this dataset. 
 
In this work, we go beyond the approach of Ref.~\cite{RGJ:2026Transparency}, where a different radial distribution is used for each single-particle state, by considering the nuclear WD to introduce a correlation between the position and momentum of the initial nucleon. The WD was introduced in Ref.~\cite{PhysRev.40.749} and called a ``probability function’’ of the simultaneous values of position and momentum of a quantum system. Of course, as remarked immediately by Wigner, the WD is not a true probability distribution but this should not hinder one to use it, many applications are reviewed in Ref.~\cite{10.1063/1.5046663}.

In the context of nuclear reactions, the WD plays a central role in kinetic transport theory~\cite{Bertsch:1988ik,Buss:2011mx}, in particular through the Boltzmann--Uehling--Uhlenbeck (BUU) equation, which the Giessen group applies to neutrino interactions in the GiBUU model~\cite{Leitner:2008ue,
Mosel:2016cwa}. While we do not solve coupled transport equations, the major difference with the GiBUU treatment is the quantum nature of the WD used in this work. In
GiBUU the ground-state WD is obtained in the LDA with position-- and momentum--dependent potentials. It is a true
classical probability distribution (quantum effects arise in the time evolution through the collision terms) and does not feature shell structure. Here, the WD is obtained from bound-state solutions of the Schr\"odinger (or Dirac) equation with central potentials~\cite{Nikolakopoulos:2025fgw}; it features quantum effects and the shell structure of nuclei. 

To incorporate this WD as a phase-space density, one has to deal with the fact that it takes on negative values. A distribution with negative regions cannot be sampled directly: its negative
parts must either be carried by negative event weights or be removed without changing
the observables of interest. This problem is not specific to neutrino interactions, and
the methods we propose to solve it can be used in other MC simulations that start from
a WD.

In this paper we propose (for the first time) methods with which realistic WDs can be used in modeling neutrino interactions in MC event generators. We have implemented in \textsc{NuWro} two algorithms for the computation of neutrino interaction cross sections using the WD. 
The first retains the signed, angle-averaged WD by sampling its positive and negative branches. The second constructs an approximate non-negative phase-space distribution for unweighted event generation. We will demonstrate that both approaches lead to very similar results.

We show that by using the WD as a phase-space density, the inclusive cross section is guaranteed to be non-negative; it is equivalent to the plane-wave impulse approximation (PWIA). The extension to PWIA calculations with realistic spectral functions is straightforward. However, our approach introduces a unique correlation between the position and momentum at which nucleons are introduced in the INC. The correlations affect FSI and the implication for the JSNS$^2$ data~\cite{JSNS2:2024bzf} will be quantified in this paper.

Notably, there are recent calculations of nuclear WD that included the effect of short-range correlations~\cite{Cosyn:2021ber}, and are obtained ab initio using variational Monte Carlo~\cite{Rocco:2026buz}. It will be interesting to see if our methods can be extended to incorporate these more general approaches in the future. 

The paper is organized as follows. Section~\ref{sec:wigner_generalities} introduces the nuclear WD, its relation to Fermi gas (FG) descriptions, and the shell-resolved input tables. Section~\ref{sec:interaction_framework} describes MC event generation, \textsc{NuWro}, the interaction framework, and the treatment of nuclear binding. Section~\ref{sec:wigner_implementation} summarizes the \textsc{NuWro} implementations; their
details are given in Appendix~\ref{app:wigner_implementation}. Section~\ref{sec:results} compares the predictions with
electron-- and neutrino--scattering data and examines their sensitivity
to position--momentum correlations.
Section~\ref{sec:summary} summarizes the results of this work and possible extensions. 

\section{Wigner distributions: generalities}
\label{sec:wigner_generalities}

The WD~\cite{PhysRev.40.749,Hillery:1983ms} is a quantum counterpart of a classical phase-space distribution. Its integrals over momentum
and over position give the exact coordinate-- and momentum--space densities, and it allows one to calculate expectation values of operators that are sums of position and momentum dependent operators.
Because of the uncertainty principle the WD can (and generally will) take on negative values, so it cannot be treated as a joint probability of position and momentum. In this work we treat the negative regions explicitly, obtaining physically meaningful results.

\subsection{Definition}
\label{subsec:definition}

We use natural units, $\hbar=c=1$. The nuclear WD for nucleons of isospin $\tau\in\{p,n\}$ is defined as
\begin{equation}
W_\tau(\mathbf r,\mathbf p)
=
\int d^3s\,
e^{-i\mathbf p\cdot\mathbf s}
\rho_\tau\!\left(
\mathbf r-\frac{\mathbf s}{2},
\mathbf r+\frac{\mathbf s}{2}
\right),
\label{eq:wigner_definition}
\end{equation}
where $\rho_\tau(\mathbf r,\mathbf r^\prime)$ is the corresponding
one-body density matrix and \(\{p,n\}\) corresponds to protons and neutrons. The WD is normalized such that

\begin{align}
\rho_\tau(\mathbf r)
&=
\int\frac{d^3p}{(2\pi)^3}\,
W_\tau(\mathbf r,\mathbf p),
\label{eq:wigner_coordinate_marginal}\\
n_\tau(\mathbf p)
&=
\int d^3r\,
W_\tau(\mathbf r,\mathbf p).
\label{eq:wigner_momentum_marginal}
\end{align}
Here, $\rho_\tau(\mathbf r)$ and $n_\tau(\mathbf p)$ are the
coordinate- and momentum-space number densities, respectively. Although the full WD depends on the cosine of the relative angle $\cos\theta_{rp} = \hat{\mathbf{r}}\cdot\hat{\mathbf{p}}$, it is more convenient to consider the angle-averaged distribution 
\begin{equation}
    w_\tau(r,p) \equiv \frac{1}{2} \int d\cos\theta_{rp} \, W_\tau(\mathbf{r}, \mathbf{p})
\end{equation}

WD is a quasiprobability distribution, that in the classical limit gives the joint probability of finding a nucleon with momentum $\mathbf{p}$ at position $\mathbf{r}$. In infinite nuclear matter, and in a LFG approach, it is equivalent to the occupation probability of a momentum state in the Fermi sea.  In a finite nuclear system, this direct interpretation is lost. We discuss these approximations in Sec.~\ref{sec:FG}.

\subsection{Wigner distribution for the Fermi gas models}
\label{sec:FG}

The FG approaches are based on non-interacting infinite nuclear matter~\cite{Moniz:1971mt}. In this case nucleons fully occupy momentum states up to a fixed Fermi momentum $k_F$.
The occupation number for nucleons of isospin $\tau$ is hence
\begin{equation}
  f_\tau(\mathbf{p}) = \Theta\bigl(k_{F,\tau} - |\mathbf{p}|\bigr) ,
  \label{eq:fg_occupation}
\end{equation}
so that the density is
\begin{equation}
  \rho_\tau = 2 \int \frac{d^3p}{(2\pi)^3}\, f_\tau(\mathbf{p})
            = \frac{k_{F,\tau}^3}{3\pi^2} ,
  \label{eq:fg_density}
\end{equation}
where the factor of two accounts for the spin degeneracy. As shown in Ref.~\cite{Nikolakopoulos:2025fgw}, the WD is then given by twice (the spin-degeneracy) the occupation number.

A finite nucleus can be approximated in the LDA~\cite{Nieves:2004wx} by considering a number density $\rho_\tau(\mathbf{r})$, where 
\begin{equation}
    N_\tau = \int d^3r \rho_\tau(\mathbf{r}),
\end{equation}
is the number of nucleons of isospin \(\tau\).  
One then defines the local Fermi momentum as
\begin{equation}
  k_{F,\tau}(\mathbf{r}) = \bigl[3\pi^2 \rho_\tau(\mathbf{r})\bigr]^{1/3},
  \label{eq:kF_local}
\end{equation}
so that the WD in the LFG is
\begin{equation}
  W^{\rm LFG}_\tau(\mathbf{r},\mathbf{p})
    = 2\,\Theta\bigl(k_{F,\tau}(\mathbf{r}) - |\mathbf{p}|\bigr).
  \label{eq:W_LFG}
\end{equation}
The global Fermi gas model (GFG), where $k_F$ is a fixed value, may  be interpreted as a nucleus represented by a ball of radius $R$ with constant density.
The radius is fixed by $\tfrac{4\pi}{3}R^3\rho_\tau = N_\tau$, and the corresponding WD is
\begin{equation}
  W^{\rm GFG}_\tau(\mathbf{r},\mathbf{p})
    = 2\,\Theta\bigl(R - r\bigr)\,
         \Theta\bigl(k_{F,\tau} - p\bigr).
  \label{eq:W_GFG}
\end{equation}

The WDs for LFG and GFG are thus non-negative phase space distributions of nucleons in the nucleus, exactly representing the joint probability of finding a nucleon with certain momentum at a certain position. Indeed, for both cases one can write
\begin{equation}
    W^{\rm FG}_\tau(\mathbf{r},\mathbf{p})
      = \rho_\tau(\mathbf{r})\, P_\tau(\mathbf{p}|\mathbf{r}) 
        \label{eq:W_FG}
      \end{equation}
where
    \begin{equation}  
    P_\tau(\mathbf{p}|\mathbf{r})
      = \frac{6\pi^2}{k_{F,\tau}^3(\mathbf{r})}\,
         \Theta\bigl(k_{F,\tau}(\mathbf{r}) - |\mathbf{p}|\bigr) 
\end{equation}
is the unit-normalized momentum distribution at fixed position. The label FG is used to denote both GFG and LFG in Eq.~\eqref{eq:W_FG} and from here on. In GFG~\eqref{eq:W_GFG}, this distribution is independent of $r$ i.e. \(k_{F,\tau}(\mathbf{r}) = k_{F,\tau}\), and there is no correlation between position and momentum. The fact that these distributions are positive follows from the starting point of non-interacting infinite nuclear matter.
In this case the momentum is a good quantum number, and of course there is no \(r\)--dependence. The \(r\)--dependence is then introduced through the LDA and in LFG, \(k_{F,\tau}(\mathbf{r})\) is calculated from a realistic \(\rho_{\tau}(\mathbf{r})\) profile -- this dependence is the only source of correlation between $\mathbf{r}$ and $\mathbf{p}$. The resulting distribution is completely classical, there is no uncertainty relation for position and momentum.

\subsection{Wigner distributions in finite nuclei}
\label{subsec:wigner_in_nuclei}

Throughout this work  $\alpha$ denotes a single-particle shell labeling the principal quantum number, total and orbital angular momentum, $\alpha\equiv(n,J,\ell)$.  We denote by $\mathcal{S}_{\tau}$ the set of
occupied shells of isospin $\tau$. In this paper we will use the WD obtained in Ref.~\cite{Nikolakopoulos:2025fgw}, we briefly outline the calculation here.
In a system with spherical symmetry, the one-body density matrix may be decomposed into states with definite angular momentum
\begin{equation}
\rho_{\tau}(\mathbf r,\mathbf r')
=
\sum_{\alpha\in\mathcal S_\tau}\sum_{m_J}
v_{\tau\alpha}^2
\left[\phi_{\tau\alpha}^{m_J}(\mathbf r)\right]^\dagger
\phi_{\tau\alpha}^{m_J}(\mathbf r'),
\label{eq:one_body_matrix}
\end{equation}
where $m_J$ is the projection of total angular momentum and \(v^2_{\tau\alpha} \leq 1\) is the occupation factor. In a (relativistic) mean-field (MF) theory, these are the occupied bound-state solutions, each with energy-eigenvalue $E_{\tau\alpha}$.
After summing over the $(2J+1)$ energy-degenerate $m_J$ states, we define
the contribution to the one-body density matrix from a single occupied
shell $\alpha$ with nucleon isospin $\tau$ as
\begin{equation}
   \rho_{\tau\alpha}(\mathbf r,\mathbf r')
=
\sum_{m_J}
v_{\tau\alpha}^2
\left[\phi_{\tau\alpha}^{m_J}(\mathbf r)\right]^\dagger
\phi_{\tau\alpha}^{m_J}(\mathbf r')
\end{equation}
and thereby define the WD for a single shell as
\begin{equation}
W_{\tau\alpha}(\mathbf r,\mathbf p)
=
\int d^3s\,
e^{-i\mathbf p\cdot\mathbf s}
\rho_{\tau\alpha}\!\left(
\mathbf r-\frac{\mathbf s}{2},
\mathbf r+\frac{\mathbf s}{2}
\right),
\label{eq:wigner_def_2}
\end{equation}

Since Eq.~\eqref{eq:wigner_def_2} is linear in the one-body density matrix, every quantity introduced in Sec.~\ref{subsec:definition} decomposes shell by shell. We denote by \(w_{\tau\alpha}(r,p)\), \(\rho_{\tau\alpha}(r)\), and \(n_{\tau\alpha}(p)\) the shell-resolved angle-averaged WD and the radial and momentum shell marginals, obtained from Eqs.~\eqref{eq:wigner_coordinate_marginal} and \eqref{eq:wigner_momentum_marginal} with \(W_{\tau}\to W_{\tau\alpha}\). Summing over the occupied shells recovers the nuclear quantities,

\begin{equation}
    X_\tau = \sum_{\alpha \in \mathcal{S}_\tau} X_{\tau\alpha}, \quad X \in \{W,w,\rho,n\}.
    \label{eq:shell_sum}
\end{equation}

The formulae for calculations of these distributions are given in the appendix of Ref.~\cite{Nikolakopoulos:2025fgw} for relativistic (four-component) and non-relativistic (two-component) wavefunctions. For use in \textsc{NuWro}, the WD of each occupied proton and neutron shell is provided as a separate two-dimensional table of $w_{\tau\alpha}(r,p)$ on an $(r,p)$ grid. The Wigner tables for $^{12}\mathrm{C}$ and $^{16}\mathrm{O}$ were calculated in the RMF~\cite{Horowitz1991} using the NL-SH parameter set of Ref.~\cite{SHARMA1993377}. Those for $^{40}\mathrm{Ar}$ and $^{56}\mathrm{Fe}$ were calculated using non-relativistic wavefunctions obtained using a Hartree-Fock calculation~\cite{Bennaceur:2005mx}
with the SLy4 functional~\cite{CHABANAT1998231}. The occupancies of the tabulated shells, listed in Table~\ref{tab:separation_energies}, follow from integrating either marginal
\begin{equation}
\begin{aligned}
N_{\tau\alpha}
&= \int d^3r \, \rho_{\tau\alpha}(r) \\
&= \int \frac{d^3p}{(2\pi)^3} \, n_{\tau\alpha}(p) \\
&= (2J+1)\,v^2_{\tau\alpha}.
\end{aligned}
\label{eq:shell_strength}
\end{equation}

\begin{figure*}
\centering\includegraphics[width=\textwidth]{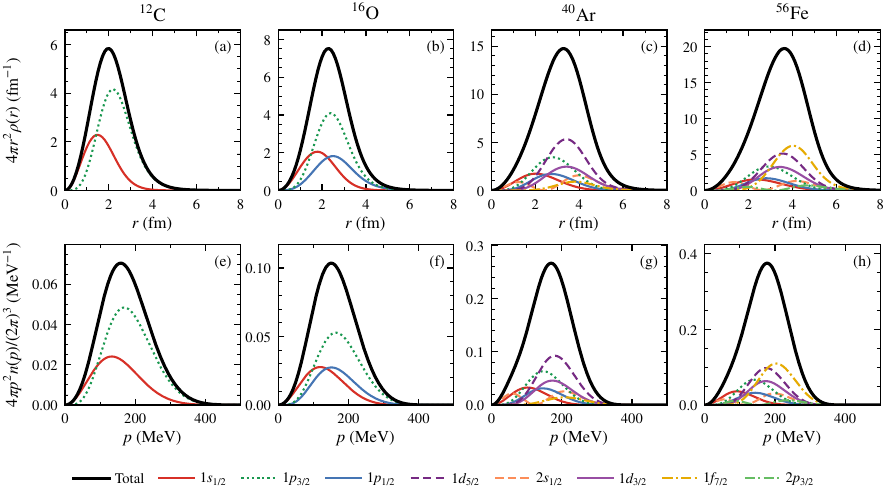}
\caption{Shell-resolved and total radial and momentum distributions obtained from the shell WDs, summed over isospin,
for $^{12}$C, $^{16}$O, $^{40}$Ar, and $^{56}$Fe. The top (bottom) panel
shows the radial (momentum) marginals.}
\label{fig:wigner_marginals_all}
\end{figure*}

For $^{12}$C and $^{16}$O, $v^2_{\tau\alpha}=1$, and for $^{40}$Ar and $^{56}$Fe the occupation factors are $v^2_{\tau\alpha}\le 1$ only for the least bound states. There are 4 shell tables for $^{12}$C, 6 for $^{16}$O, 13 for $^{40}$Ar, and 15 for $^{56}$Fe.  Each table is defined on a regular $100\times300$ grid in radial position and momentum, respectively. The radial grid contains 100 points from $r=0.01$ to $9.91$~fm with a spacing of $\Delta r=0.1$~fm, while the momentum grid contains
300 points from $p=0$ to $598$~MeV with a spacing of $\Delta p=2$~MeV. The shell-resolved radial and momentum distributions obtained from these tables are displayed in Fig.~\ref{fig:wigner_marginals_all}.

An important feature of these tables is the presence of regions with negative values (see Fig.~\ref{fig:C12-proton-Wigner-branches} for the $^{12}$C proton shells). This feature follows from the quantum nature of the states, i.e. the uncertainty principle~\cite{Hillery:1983ms,Kenfack:2004}. 
Specifically, Eq.~(\ref{eq:wigner_def_2}) probes the overlap of the wavefunctions at different positions, and the Fourier transform will then introduce oscillations at a momentum scale associated with natural length scales present in the coordinate space wavefunctions~\cite{Prakash:1981ilg}.

To show why and where $w_{\tau\alpha}(r,p)$ takes on negative values consider the appendix of Ref.~\cite{Nikolakopoulos:2025fgw}, which shows that $w_{\tau\alpha}(r,p)$ evaluated on the boundary, i.e. for $p=0$ (or, $r=0$), is determined by the Fourier transform of the radial (or, momentum) distribution of the state multiplied by $(-1)^\ell$. This implies that $w_{\tau\alpha}(0,0) = v_{\tau\alpha}^2\,8(-1)^\ell(2J+1)$ for non-relativistic wavefunctions\footnote{For relativistic wavefunctions, the value is slightly smaller than this bound because there is a difference between scalar and vector densities.}, i.e. negative for odd $\ell$. Furthermore, this implies that one expects a zero $w_{\tau\alpha}(r=0, p=p_\ast)=0$ for some $p_\ast$ that is inversely proportional to the size of the state. (As a concrete example, in a simple Gaussian model for a $p$-state\footnote{Consider $\rho_{\tau\alpha}(r)\propto r^2 e^{-2ar^2}$, then $r_{\max}=1/\sqrt{2a}$,
and the Fourier transform of the density has a zero $\tilde\rho_{\tau\alpha}(k_\ast)=0$
for $k_\ast=\sqrt{6}/r_{\max}$, the extra factor two follows from
$w_{\tau\alpha}(0,p)\propto\tilde\rho_{\tau\alpha}(2p)$ as shown in Ref.~\cite{Nikolakopoulos:2025fgw}.} one obtains $p_\ast = \sqrt{6}/(2r_{\rm max})$, where $r_{\rm max}$ is the peak of the density $\rho_{\tau\alpha}(r)$ of the state.) Then, since the WDs are smooth, these features are not restricted to the boundaries. These considerations imply that, if we consider a $p$-shell state for example, $w_{\tau\alpha}(r,p)$ is negative for small $(r,p)$ and turns positive at some boundary determined by the widths of the distributions in coordinate and momentum space. Although the regions with negative values of $w_{\tau\alpha}(r,p)$  can be sizable, their integrated contribution remains always relatively small. To quantify the negative content of the input WDs we define the negative fraction as

\begin{equation}
f^\tau_- =
\frac{
\displaystyle
\sum_{\alpha\in\mathcal{S}_\tau}
\int dr\,r^2
\int dp\,p^2
\max[-w_{\tau\alpha}(r,p),0]
}{
\displaystyle
\sum_{\alpha\in\mathcal{S}_\tau}
\int dr\,r^2
\int dp\,p^2
|w_{\tau\alpha}(r,p)|
}.
\label{eq:neg_fraction}
\end{equation}
To compare the spatial location of positive and negative contributions inside the nucleus, we define the mean radius of a signed contribution as

\begin{equation}
\langle r\rangle^\tau_\pm =
\frac{
\displaystyle
\sum_{\alpha\in\mathcal{S}_\tau}
\int dr\,r^3
\int dp\,p^2
\max[\pm w_{\tau\alpha}(r,p),0]
}{
\displaystyle
\sum_{\alpha\in\mathcal{S}_\tau}
\int dr\,r^2
\int dp\,p^2
\max[\pm w_{\tau\alpha}(r,p),0]
}.
\end{equation}

The resulting negative fractions and mean radial positions of the signed contributions for the nuclei considered in this work are summarized in Table~\ref{tab:wigner_negative_fraction}. \(f^{\tau}_{-}\) increases with nuclear mass, reflecting the increasingly oscillatory phase-space structure of the shell-resolved WDs in heavier nuclei. Negative branches are more concentrated at smaller radii than the positive branches.

\begin{table}[b]
\caption{
Integrated negative fractions and mean radii of the positive and
negative branches of the shell-resolved proton and neutron WDs. \(f_{-}^{\tau}\) are in \(\%\) and \(\langle r\rangle_\pm^\tau\) are in fm.
}
\label{tab:wigner_negative_fraction}
\begin{ruledtabular}
\begin{tabular}{lcccccc}
Target
& $f_-^p$ 
& $f_-^n$
& $\langle r\rangle_-^p$ 
& $\langle r\rangle_-^n$ 
& $\langle r\rangle_+^p$ 
& $\langle r\rangle_+^n$ 
\\
\hline
$^{12}\mathrm C$  & 3.7  & 3.7  & 0.96 & 0.95 & 2.11 & 2.09 \\
$^{16}\mathrm O$  & 4.3  & 4.3  & 1.05 & 1.04 & 2.36 & 2.33 \\
$^{40}\mathrm{Ar}$ & 8.8 & 9.3  & 1.83 & 1.87 & 3.03 & 3.10 \\
$^{56}\mathrm{Fe}$ & 10.0 & 10.7 & 2.06 & 2.14 & 3.33 & 3.35 \\
\end{tabular}
\end{ruledtabular}
\end{table}

\section{Monte Carlo event generators and neutrino interactions}
\label{sec:interaction_framework}

\subsection{Monte Carlo event generators}

In general, MC generators~\cite{Andreopoulos:2009rq,Hayato:2021heg,Juszczak:2005zs,Buss:2011mx,Isaacson:2022cwh} compute neutrino scattering cross sections by integrating over the available phase
space with the MC algorithm~\cite{Mosel:2019vhx}. In the computation, points of the final-state phase space are sampled; each of them
corresponds to a possible configuration of the final-state particles.

In an ideal situation, if all the exclusive interaction cross sections are known, for each sampling we calculate a product of the differential cross section and the volume of the available phase space obtaining event weight $\omega_j$. In practice, exclusive final states are modeled with various approximations, but the notion of event weight is valid. For a sample of \(N_{\mathrm{gen}}\) generated phase-space points, the average weight is the estimator of the total cross section
\begin{equation}
\sigma
\simeq
\frac{1}{N_{\mathrm{gen}}}
\sum_{j=1}^{N_{\mathrm{gen}}} \omega_j.
\label{eq:total_sigma}
\end{equation}

\subsection{NuWro}

\textsc{NuWro} has been serving the neutrino community as a reliable MC event generator of lepton-nucleus interactions for more than 20 years. For neutrino--nucleus scattering, \textsc{NuWro} offers several reaction mechanisms such as quasielastic (QE)~\cite{LlewellynSmith:1971uhs}, hyperon production (HYP), resonance excitation (RES), deep-inelastic scattering (DIS), and neutrino--electron scattering (EL). Meson-exchange currents,
and coherent pion production (COH) are also considered for interactions on nuclear targets. After the primary vertex interaction, hadronic reinteractions are modeled in \textsc{NuWro}'s INC model~\cite{Golan:2012wx, Niewczas:2019fro}. 

In \textsc{NuWro}, one can choose from different descriptions of the initial-state bound target nucleons. For QE scattering, \textsc{NuWro} has options to choose from GFG or LFG models, momentum-- and density--dependent potentials~\cite{Juszczak:2003zw}, and the spectral-function framework~\cite{Banerjee:2023hub,Ankowski:2025umq}. The present work adds a shell-resolved WD description to the QE interaction framework. 

In \textsc{NuWro}, MC event generation can be performed either in weighted or unweighted mode~\cite{Juszczak:2009qa}. In weighted generation, the sampled phase-space points are retained together with their generally different event weights, and physical distributions are obtained from weighted sums. In unweighted generation, events are selected with the acceptance--rejection algorithm and have equal weight.

\subsection{Cross section formula}
\label{subsec:lfg_cross_section}

We first write the phase-space structure of the expression for the CCQE cross section in the FG description~\cite{Smith:1972xh}.
\begin{align}
\frac{d^2\sigma_{\rm FG}}{dE_\ell d\Omega_\ell}
={}&
\mathcal{N}
\sum_{\tau}
\int d^3r\,\rho_\tau(\mathbf r)
\int \frac{d^3p}{(2\pi)^3}\,
P_\tau(\mathbf p|\mathbf r)\,
\mathcal{K}_\tau(\mathbf p,\mathbf q)
\nonumber\\
&\times
\Theta^{\rm PB}\!\left(
|\mathbf p+\mathbf q|-k_{F,\tau'}
\right),
\label{eq:lfg_cross_section}
\end{align}

where \(\Theta^{\rm PB}\) implements Pauli-blocking by requiring the momentum of the outgoing nucleon with isospin \(\tau'\) to lie above the global Fermi momentum. The coefficient \(\mathcal{N}\) and elementary kernel \(\mathcal{K}_\tau(\mathbf p,\mathbf q)\) are defined as

\begin{equation}
 \mathcal{N} =   \frac{G_F^2\cos^2\theta_C}{(2\pi)^2}\frac{|\mathbf k'|E_\ell}{E_\nu}, \, \mathcal{K}_\tau(\mathbf p,\mathbf q) = \frac{M}{E_p}
\mathcal{L}_{\mu\nu}
\mathcal{W}_\tau^{\mu\nu}(\mathbf p,\mathbf q). 
\end{equation}

Other notations are explained in Refs.~\cite{LlewellynSmith:1971uhs,Formaggio:2012cpf}. The FG cross
section can be read as an integration of the elementary neutrino--nucleon cross section over that occupied phase space, with
$W^{\rm FG}_\tau = \rho_\tau(\mathbf{r})P_\tau(\mathbf{p}|\mathbf{r})$ acting as the phase-space weight. Replacing the FG occupation function by the finite-nucleus WD, the cross section reads

\begin{align}
\frac{d^2\sigma_{\rm WD}}
     {dE_\ell d\Omega_\ell}
={}&
\mathcal{N}
\sum_{\tau}
\int \int d^3r\frac{d^3p}{(2\pi)^3}\,
W_\tau(\mathbf r,\mathbf p)\,
\mathcal{K}_\tau(\mathbf p,\mathbf q)
\nonumber\\
&\times
\Theta^{\rm PB}\!\left(
|\mathbf p+\mathbf q|-k_{F,\tau'}
\right).
\label{eq:wigner_cross_section}
\end{align}

It is important to note that, while generally $W_\tau(\mathbf{r},\mathbf{p})$ can take on negative values, the cross section obtained from Eq.~(\ref{eq:wigner_cross_section}) is guaranteed to be positive. Indeed, Eq.~(\ref{eq:wigner_momentum_marginal}) allows us to perform the integral over $\mathbf{r}$, so that for any fixed shell $\alpha$ of nucleon with isospin $\tau$ one obtains

\begin{align}
\frac{d^2\sigma^{\tau\alpha}_{\rm WD}}
     {dE_\ell d\Omega_\ell}
={}&
\mathcal{N}
\int \frac{d^3p}{(2\pi)^3}\,
n_{\tau\alpha}(p)\,
\mathcal{K}_\tau(\mathbf p,\mathbf q)
\nonumber\\
&\times
\Theta^{\rm PB}\!\left(
|\mathbf p+\mathbf q|-k_{F,\tau'}
\right),
\label{eq:PWIA}
\end{align}
with \(n_{\tau\alpha}\) being the momentum shell marginal introduced in Sec.~\ref{subsec:wigner_in_nuclei}. Equation~\eqref{eq:PWIA} is exactly the result obtained in the PWIA. In summary, starting from the FG prescription but with a realistic shell-separated WD, we obtain the PWIA cross section. The novel feature that arises in this case is that a non-trivial correlation between the initial nucleon momentum $p$ and the position of the nucleon $r$ is now introduced. This correlation will affect the FSI effects. The results go further than those recently presented in Ref.~\cite{RGJ:2026Transparency}, in which a correlation between the specific single particle state $\alpha$ and the position at which the nucleon is introduced in the INC is considered, but without a correlation between $r$ and \(p\). That is, within the framework presented here, the result of Ref.~\cite{RGJ:2026Transparency} can be seen as an approximation of the WD as the product of radial and momentum distributions

\begin{equation}
    w_{\tau\alpha}(r,p) \approx \frac{\rho_{\tau\alpha}(r) \,n_{\tau\alpha}(p)}{N_{\tau\alpha}}.
\end{equation}

Finally, the extension to PWIA results with realistic SFs is obvious, at least for the MF contribution.
Assume that the MF SF is given by
\begin{equation}
    S_\tau(p,E) = \sum_{\alpha\in\mathcal{S}_\tau}  n_{\tau\alpha}(p)\,\mathcal{D}_{\tau\alpha}(E),
    \label{eq:sf}
\end{equation}
with $\mathcal{D}_{\tau\alpha}(E)$ being the removal-energy distribution for the state $\alpha$ normalized as

\begin{equation}
    \int dE\, \mathcal{D}_{\tau\alpha}(E) = 1.
\end{equation}

Then, by inspection of Eqs.~\eqref{eq:wigner_cross_section} and~\eqref{eq:PWIA}, for any event with initial nucleon momentum $p$, one should distribute the position of the nucleon according to the following conditional radial distribution function 

\begin{equation}
\label{eq:conditional_wigner_PWIA}
 P_{\tau\alpha}(r\mid p) = \frac{w_{\tau\alpha}(r,p)}{n_{\tau\alpha}(p)}.
\end{equation}

One could explore the same ansatz in distorted wave calculations, but one should include in principle the proper overlap with the final-state WD to guarantee strictly positive results.
This will be explored in future work.

\subsection{Nuclear binding}
\label{sec:binding}

Equation~\eqref{eq:wigner_cross_section} is not yet complete, because it does not specify the energy of the struck nucleon. A bound nucleon
is off shell: its energy differs from the free value $E_p = \sqrt{\mathbf{p}^2 + m_\tau^2}$. We write it as
\begin{equation}
  p^0 = E_p - B ,
  \label{eq:p0_B}
\end{equation}

where $B\geq 0$ is the binding energy, so that energy conservation at the primary vertex reads
\begin{equation}
  E_p + E_\nu - B = E_\ell + E_{p'} .
  \label{eq:IA_balance}
\end{equation}

In general, $B$ is not constant and depends on the position of the nucleon inside the nucleus and on its momentum. For an off-shell nucleon, the hadronic current entering  $\mathcal{K}_\tau$ is not uniquely defined, and a prescription is needed~\cite{DEFOREST1983232}. In \textsc{NuWro}, the outgoing lepton and nucleon are generated on shell from four-momentum
conservation with the off-shell initial nucleon $(p^0,\mathbf{p})$. The
elementary cross section is the free-nucleon one~\cite{LlewellynSmith:1971uhs},
evaluated at the resulting $q^2$ and at the neutrino energy in the rest frame of
an on-shell nucleon with momentum $\mathbf{p}$. Binding enters through
the energy balance in Eq.~\eqref{eq:IA_balance}.

\subsubsection{Fermi gas models}

In the standard GFG, $B$ is a constant, treated as a phenomenological parameter \cite{Moniz:1971mt}. In \textsc{NuWro}, $B^{\rm GFG}$ is a nucleus-dependent input parameter. For LFG in \textsc{NuWro}, binding is represented through an effective mean-field potential, containing both the Fermi kinetic energy and an estimate of the energy needed to be transferred to remove a nucleon from the system. The local Fermi kinetic energy is, 

\begin{equation}
\label{eq:TF}
T_{F,\tau}(\mathbf{r}) = \sqrt{k^2_{F,\tau}(\mathbf{r})+m^2_\tau}-m_\tau .
\end{equation}

and the LFG binding energy reads

\begin{equation}
\label{eq:B_FG}
B^{\rm LFG}_\tau(\mathbf{r}) = T_{F,\tau}(\mathbf{r}) + \mathcal{E} .
\end{equation}
Currently, in \textsc{NuWro} $\mathcal{E}$ is taken to be $7$~MeV.

\subsubsection{Wigner distribution}

\begin{figure}[t]
    \centering
\includegraphics[width=0.9\linewidth]{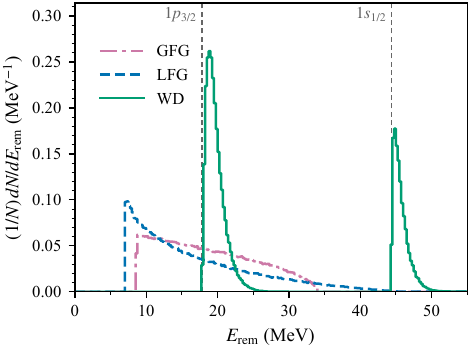}
    \caption{Shape of removal-energy $E_{\rm rem}$ distributions corresponding to Eqs.~\eqref{eq:B_FG} and \eqref{eq:B_WD} for $^{12}$C in the GFG (with \(B=34\) MeV), LFG, and WD descriptions for 1 GeV \(\nu_\mu\) QE interaction. Dashed vertical lines mark \(S_{n\alpha}\) of the $1p_{3/2}$ and $1s_{1/2}$ shells (see Table~\ref{tab:separation_energies}). For $\bar\nu_\mu$
scattering the struck nucleons are protons, and the WD peaks start at
$S_{p\alpha}=14.5$ and $40.5$~MeV. 
}
    \label{fig:binding_energy}
\end{figure}

The WD description retains the shell $\alpha$ from which the nucleon is
removed, so $B$ follows from the energy balance of the reaction. For a nucleus at rest, removing a nucleon with momentum
$\mathbf{p}$ from shell $\alpha$ leaves the residual system with momentum
$-\mathbf{p}$ and mass $M_{A-1} = M_A - m_\tau + S_{\tau\alpha}$, where
$S_{\tau\alpha}$ is the separation energy of the shell
(see Table~\ref{tab:separation_energies}). With the residual system on its mass shell (i.e. neglecting excitation energy), exact energy conservation for CCQE events with single nucleon knock-out reads
\begin{equation}
  E_\nu + M_A = E_\ell + E_{p'} + \sqrt{\mathbf{p}^2 + M_{A-1}^2} .
  \label{eq:exact_balance}
\end{equation} 

This coincides with the energy balance of Eqs.~\eqref{eq:p0_B}
and~\eqref{eq:IA_balance}, $E_\nu + p^0 = E_\ell + E_{p'}$, if
\begin{equation}
  p^0 = M_A - \sqrt{\mathbf{p}^2 + M_{A-1}^2}
      = m_\tau - S_{\tau\alpha} - T_{A-1} ,
  \label{eq:p0_WD}
\end{equation}
where 
\begin{equation}
T_{A-1} = \sqrt{\mathbf{p}^2 + M_{A-1}^2} - M_{A-1}
\label{eq:recoil}
\end{equation}
is the recoil
energy of the residual system. The binding energy in
Eq.~\eqref{eq:IA_balance} is then
\begin{equation}
  B^{\rm WD}_{\tau\alpha}(\mathbf{p}) = E_p - p^0
    = S_{\tau\alpha} + T_{A-1} + T_\tau ,
  \label{eq:B_WD}
\end{equation}
with $T_\tau = E_p - m_{\tau}$. 

Another way to express the binding energy is removal energy defined as 
\begin{equation}
E_{\mathrm{rem}} = B - T_\tau = S_{\tau\alpha} + T_{A-1}.
\label{eq:Erem}
\end{equation}

Figure~\ref{fig:binding_energy} compares the removal-energy distributions for GFG, LFG and WD for $^{12}$C. The shell structures of $^{12}$C are clearly
visible in the WD case, where the two peaks correspond to the \(1p_{3/2}\) and \(1s_{1/2}\) shells at 17.8 and 44.4~MeV respectively, broadened only by the nuclear recoil energy.

\section{Implementation in NuWro}
\label{sec:wigner_implementation}

The WD framework is implemented in \textsc{NuWro} to introduce a joint description of the position and momentum of the struck nucleon. In the LFG-based sampling, the interaction position and initial-nucleon momentum are not completely independent, since the locally allowed momentum range is determined by the position-dependent Fermi momentum. But it does not contain any information of the nuclear shell structure. In the present implementation, these quantities are connected through \(w_{\tau\alpha}(r,p)\).

The implementation of WD in an MC event generator is not straightforward because, unlike an ordinary probability distribution, the WD is not positive definite. Its negative regions cannot be interpreted as probabilities and prevent direct sampling. To incorporate the full Wigner phase-space information in \textsc{NuWro}, we consider two alternative approaches. 

In the weighted mode, the original signed WD is retained by sampling its positive and negative components separately and incorporating the corresponding sign and normalization correction into the event weight. In the unweighted mode, we instead construct a non-negative coarse-grained distribution that can be sampled probabilistically while preserving the original shell momentum distributions exactly. 

In both modes events are generated as follows. The isospin \(\tau\) of the struck nucleon is fixed by the type of interaction. The shell \(\alpha\) (in the weighted mode together with the branch) and the radial position \(r\) are sampled first, and the momentum magnitude \(p\) is then sampled from the conditional distribution at that \(r\). The directions of \(\mathbf{r}\) and \(\mathbf{p}\) are sampled isotropically. The binding energy is evaluated from Eq.~\eqref{eq:B_WD}, and the outgoing lepton and nucleon are generated as described in Sec.~\ref{subsec:lfg_cross_section}. The nucleon then enters the INC at the sampled position \(\mathbf{r}\). Detailed discussions of the sampling procedures are presented in Appendix~\ref{app:wigner_implementation}. All results of Sec.~\ref{sec:results} are
obtained in the weighted mode; the unweighted mode is used as a check in Sec.~\ref{subsec:elec} and Appendix~\ref{app:performance}.

\section{Results}
\label{sec:results}

\begin{figure*}
\centering\includegraphics[width=\textwidth]{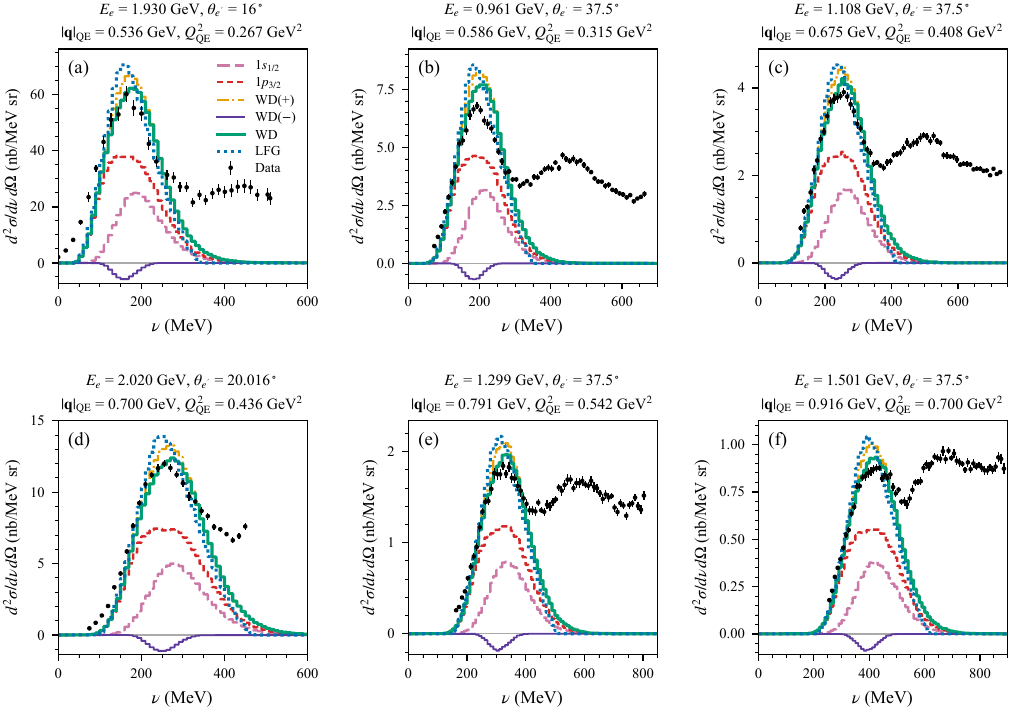}
\caption{Comparison of inclusive \(d^{2}\sigma/d\nu d\Omega\) QE cross sections predicted with WD (solid green) and LFG (dotted blue) models for $^{12}\mathrm{C}(e,e')$ measurements. WD(+) (yellow) and WD(-) (purple, drawn below zero) are the contributions of
 events sampled from the positive and negative branches of the WD in the
 weighted mode. As for the shell decomposition of the WD result: the $1p_{3/2}$ contribution is shown in red and the $1s_{1/2}$ contribution in pink. The panels are labeled according to beam
energy \(E_e\), scattering angle \(\theta_e'\), and values of \(|\mathbf{q}|\) and \(Q^2\) at the QE peak. The experimental data (black points with vertical error bars) are from Refs.~\cite{Bagdasaryan:1988hp} [panel (a)], \cite{Sealock:1989nx} [panels (b), (c), (e), (f)], and \cite{Day:1993md} [panel (d)]. }
    \label{fig:electron_scattering}
\end{figure*}

\subsection{Comparison to inclusive electron-scattering data}
\label{subsec:elec}

Figure~\ref{fig:electron_scattering} compares the WD and LFG results with inclusive $^{12}\mathrm{C}(e,e')$ data~\cite{Bagdasaryan:1988hp,Sealock:1989nx,Day:1993md,Benhar:2006er} for six different kinematic settings. \(\nu\) is the energy transfer and \(\Omega\) is the solid angle corresponding to the outgoing electron. 
Non-QE channels cannot currently be run in \textsc{NuWro}'s electron mode, so we test only the QE prediction.

The WD predictions in Fig.~\ref{fig:electron_scattering} are obtained in the weighted mode, see  Appendix~\ref{app:weighted_wigner}\footnote{For each kinematic setting we generate $2\times10^{8}$ events. The scattered electron is accepted in a window $|\cos\theta - \cos\theta_e| \le 0.001$ around the nominal angle $\theta_e$. The \textsc{NuWro} predictions in $\nu$ are in 10 MeV bins.}. Accordingly, the cross section splits into the
contributions of events sampled from the positive and the negative
branch,
\begin{equation}
  \frac{d^2\sigma_{\rm WD}}{d\nu\, d\Omega}
  = \frac{d^2\sigma^{(+)}_{\rm WD}}{d\nu\, d\Omega}
  - \frac{d^2\sigma^{(-)}_{\rm WD}}{d\nu\, d\Omega},
  \label{eq:xsec_pm}
\end{equation}
in direct correspondence with Eq.~\eqref{eq:cross_section_decomposition}. As expected, the positive contribution fully absorbs the negative one, and one obtains physically meaningful positive results. The negative contribution, about \(3.7\%\) (see Table~\ref{tab:wigner_negative_fraction}), comes almost
entirely from the $1p_{3/2}$ shell, where it is about 5.5\% of the positive
branch, while for $1s_{1/2}$ it stays below 0.1\%. As for the distribution in $\nu$, the negative contribution is concentrated around the QE peak, with a width about 2.5 times smaller than
that of the total, so that at its maximum it reaches about 9\% of the peak height. 

As a cross-check of the two implementations, we repeated the calculation in the unweighted mode with the non-negative coarse grained distribution of Appendix~\ref{app:cg_wigner}. The integrated cross sections agree with the total weighted results with precision better than 1\%, consistent with MC statistical fluctuations.

By construction, see Sec.~\ref{subsec:lfg_cross_section}, these results are exactly those obtained in the PWIA.
For an inclusive observable the radial integration in
Eq.~\eqref{eq:wigner_cross_section} can be carried out first, and the WD cross section reduces to Eq.~\eqref{eq:PWIA}.
Together with the treatment of binding energy of Eq.~(\ref{eq:B_WD}), this is exactly the PWIA with the spectral function of
Eq.~\eqref{eq:sf}, in which each shell contributes at a fixed removal-energy, $\mathcal{D}_{\tau\alpha}(E) = \delta(E-S_{\tau \alpha})$.

While the results are equivalent to the PWIA, the main feature of using the WD is that position-momentum correlations are introduced, which determine where the nucleon is placed in the INC and hence will affect results sensitive to FSI modeled using the INC. This is studied in the following subsections. Here we comment on the extension to more realistic spectral functions. 

The main discrepancies of our PWIA results with respect to the calculations of Refs.~\cite{Rocco:2018mwt, Ankowski:2014yfa, Rocco:2019gfb, Lovato:2023raf} are well understood~\cite{VanOrden:2019krz}.
The RMF single-particle states are treated as fully occupied ($v^2_{\tau\alpha}=1$) and have a fixed excitation energy.
In reality, the occupation numbers are quenched, and the spectral strength may be fragmented and smeared out over a significant energy region~\cite{BENHAR1994493, VanOrden:2019krz, BOFFI19931, JLabE91013:2003gdp, Ankowski:2024ntv}, due to short-- and long--range correlations~\cite{Barbieri:2009ej, Dickhoff:2004xx}.

Including these effects explicitly is non-trivial~\cite{Lovato:2016gkq,Niewczas:2026xox,Casale:2025zac}, but for (semi-)inclusive interactions, models based on the PWIA with effective or microscopic spectral functions are quite successful~\cite{Rocco:2019gfb, Lovato:2023raf}, and it would be interesting to apply our approach to the WDs obtained in Refs.~\cite{Rocco:2026buz, Cosyn:2021ber}.
Lastly, it remains crucial to include elastic FSI~\footnote{Following Ref.~\cite{RGJ:2026Transparency}, the ``elastic’’ FSI are those in which the outgoing nucleon does not knock out additional nucleons or exchanges energy with the residual system; the latter are ``inelastic’’ FSI which are treated in the INC but do not affect the inclusive cross section. }, which are not taken into account in the PWIA. These lead mainly to a shift and smearing of the inclusive cross section which may be included approximately~\cite{Ankowski:2014yfa, PhysRevC.44.2328}, or by treating the outgoing nucleon as a scattering state in a potential~\cite{Gonzalez-Jimenez:2019qhq, Franco-Munoz:2022jcl, RGJ:2026Transparency, Horikawa:1980cg, Meucci:2003uy}. 

As argued in Sec.~\ref{subsec:lfg_cross_section}, the WD treatment extends very easily to an effective treatment that includes reduced occupation numbers and energy smearing of single particle states. The inclusion of FSI and a treatment of correlations in a consistent manner remains to be established however. These will be explored in future work.

\subsection{Sensitivity to position--momentum correlations}
\label{sec:rp_corr}

The comparison with electron-scattering data in
Sec.~\ref{subsec:elec} provides a sanity test for the correctness of the algorithm to deal with the negative branch of WDs.  In this subsection we demonstrate an impact of the position--momentum correlations on physical observables, by comparing the original
shell-resolved WD with two factorized constructions in which increasing amounts
of correlation information are removed.

\subsubsection{Factorized Wigner distributions}
\label{subsubsec:factorized_wigner}

We consider three descriptions of the initial state. In the first, denoted CORR, the original shell-resolved WD is used directly. The second, denoted shell-FACT, removes the position--momentum correlation inside each shell while preserving both of its marginals,

\begin{equation}
w_{\tau\alpha}^{\rm shell\text{-}FACT}(r,p)
= \frac{\rho_{\tau\alpha}(r)\, n_{\tau\alpha}(p)}{N_{\tau\alpha}}.
\label{eq:wigner_shell_factorized}
\end{equation}

The information removed is the correlation between $r$ and $p$ within a given
shell, this is the construction of Ref.~\cite{RGJ:2026Transparency}. The third construction, denoted totR-FACT, additionally removes the shell dependence of the spatial distribution. We define it as

\begin{equation}
w_{\tau\alpha}^{\rm totR\text{-}FACT}(r,p)
= \frac{\rho_\tau(r)}{N_\tau}\, n_{\tau\alpha}(p).
\label{eq:wigner_totr_factorized}
\end{equation}

\begin{table}[b]
\caption{WD descriptions used to test the sensitivity to position--momentum correlations.
}
\label{tab:wigner_facts}
\begin{ruledtabular}
\begin{tabular}{lll}
Description & WD & Information retained
\\
\hline
CORR &
$w_{\tau\alpha}$ &
Full shell-dependent $r$--$p$ structure
\\[4pt]

shell-FACT &
$\displaystyle
w_{\tau\alpha}^{\rm shell\text{-}FACT}$
&
Shell-dependent $\rho_{\tau\alpha}$ and $n_{\tau\alpha}$
\\[10pt]

totR-FACT &
$\displaystyle
w_{\tau\alpha}^{\rm totR\text{-}FACT}$
&
Shell-dependent $n_{\tau\alpha}$ only \\
\end{tabular}
\end{ruledtabular}
\end{table}

This removes both the intra-shell $r$--$p$ correlation and the dependence of the spatial distribution on the shell from which the nucleon is removed. Summed over the shells, it gives $w_\tau^{\text{totR-FACT}}(r,p)=\rho_\tau(r)\,n_\tau(p)/N_\tau$: the interaction point is independent of both the momentum and the shell, which is the uncorrelated limit. This approach is typical in neutrino event generators. The shell label is kept only to assign the binding energy: for all three descriptions it follows Eq.~\eqref{eq:B_WD}, so that they share the same primary-vertex kinematics.

Because \(\mathcal{K_\tau}\) and Pauli blocking used for all three do not depend on $r$, Eq.~\eqref{eq:PWIA} applies to all
three descriptions: at the primary vertex they give identical distributions of the initial-nucleon momentum, of the removal-energy, and of the lepton kinematics. They differ only in the exclusive picture through the INC. Within the INC, \textsc{NuWro} still uses LFG distribution for the secondary nucleons.

\subsubsection{JSNS\texorpdfstring{$^2$}{2} missing-energy measurement}

\begin{figure}[!t]
    \centering
\includegraphics[width=\linewidth]{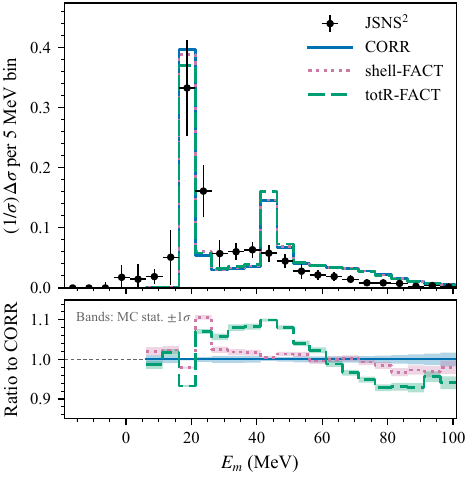}
    \caption{Missing-energy distribution for the JSNS$^2$ measurement compared with the three Wigner descriptions of Table~\ref{tab:wigner_facts}. The lower panel shows the ratios of the shell-FACT and totR-FACT predictions to the CORR result. All bins are 5~MeV wide; the $1p_{3/2}$ peak bin covers $16.2<E_m<21.2$~MeV.}
    \label{fig:kdar}
\end{figure}

The sensitivity of the three Wigner descriptions as summarized in Table~\ref{tab:wigner_facts} is studied using recent JSNS$^2$ measurement of charged-current interactions of mono-energetic \(\nu_\mu\) of the energy 235.53 MeV on \(^{12}\)C~\cite{JSNS2:2024bzf}. Due to limitations discussed in Sec.~\ref{subsec:elec} we cannot expect to reproduce the experimental results. However, the JSNS$^2$ measurement allows for a clear demonstration of the role played by position-momentum correlations, and the shell-dependence.
JSNS$^2$ reconstructs the visible energy of each event from the scintillation light
and provides the background-subtracted distribution, unfolded to the visible energy:

\begin{equation}
  E_{\rm vis} = T_\mu + \sum_p T_p ,
  \label{eq:Evis}
\end{equation}
where the sum runs over all protons leaving the nucleus, with no threshold on their
kinetic energy. Neutrons, the recoil of the residual nucleus, and de-excitation
photons are not taken into account. The result is presented in terms of the missing-energy defined as: 

\begin{equation}
\begin{aligned}
  E_m &= E_{\nu_\mu} - m_\mu + (m_n - m_p) - E_{\rm vis} \\
      &= 131.17\,{\rm MeV} - E_{\rm vis}.
\end{aligned}
\label{eq:Em_jsns}
\end{equation}

In our simulation we evaluate the quantity defined in Eq.~\eqref{eq:Em_jsns} for every event, summing up contributions from all protons in the final state after the INC. 

We notice that for QE events in which the proton leaves the nucleus without rescattering, the energy balance of Eqs.~\eqref{eq:p0_B}, \eqref{eq:IA_balance} and \eqref{eq:p0_WD} gives $\nu-T_p=S_{n\alpha}+T_{A-1}-(m_n-m_p)$, so that, with the term $m_n-m_p$ in Eq.~\eqref{eq:Em_jsns}, $E_m=S_{n\alpha}+T_{A-1}$. This is equal to the distribution of the removal energy $E_{\rm rem}$ shown in 
Fig.~\ref{fig:binding_energy}: these events yield two peaks, which start at 17.8 and 44.4~MeV for the
$1p_{3/2}$ and $1s_{1/2}$ shells respectively and are broadened only by the recoil energy.

We observe, as in Ref.~\cite{Franco-Patino:2026rwy}, that counting the recoil energy of the residual nucleus (1.9~MeV on average) as contributing to $E_{\rm vis}$ would change the
shape of the $E_m$ distribution considerably: it removes the recoil broadening of
the shell peaks, so that the bin above the $1p_{3/2}$ peak loses about 60\% of its events. 

\subsubsection{Role of position-momentum correlations}

Figure~\ref{fig:kdar} compares the JSNS$^2$ missing-energy distribution with the predictions obtained using the Wigner descriptions of Table~\ref{tab:wigner_facts}. The \textsc{NuWro} sample contains QE events
and a small MEC contribution of about 2.9\%, using the model of Ref.~\cite{Prasad:2024gnv}. The peak at lower $E_m$ is dominated by
nucleons removed from the $1p_{3/2}$ shell, and the one at higher $E_m$ by the
$1s_{1/2}$ shell. The few events below the $1p_{3/2}$ peak come solely from MEC. 

The $E_m$ distribution is sensitive to the initial state through two effects. The main one is the shell dependence of the interaction point. The nucleons knocked out from the $1s_{1/2}$ state originate from deeper in the nucleus than those from the $1p_{3/2}$. This effect is captured in the approximated shell-FACT approach and the full WD (CORR). A proton knocked out from deeper in the nucleus traverses more nuclear matter and rescatters more often. Since $E_m$ separates the two shells, when the shell dependence of the interaction point is removed as in totR-FACT, the relative height of the two peaks changes.
To understand this quantitatively the average radii at which nucleons are introduced in the INC for the different approaches are listed in Table~\ref{tab:wigner_kdar_summary}. The lower panel of Fig.~\ref{fig:kdar} shows how this radial information propagates to the $E_m$ distribution. In totR-FACT, where the shell-dependence is completely neglected, the $1p_{3/2}$ nucleons are moved inward and the $1s_{1/2}$ nucleons outward. The protons from $1p_{3/2}$ therefore rescatter more: the peak decreases by 7\%, and the strength moves to the region between the peaks.
The protons from $1s_{1/2}$ rescatter less: the peak increases by 10\%, and the high-$E_m$ tail decreases.

\begin{figure}[!t]
    \centering
\includegraphics[width=\linewidth]{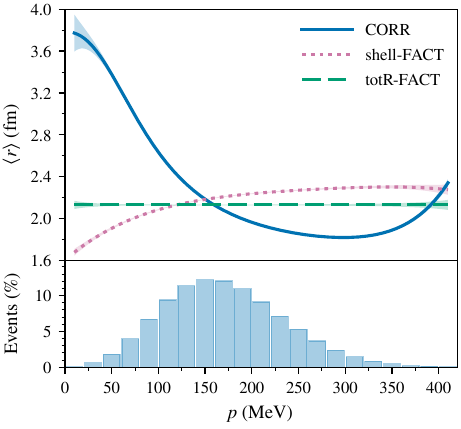}
\caption{Mean distance of the interaction point from the nuclear center as a
function of the initial-nucleon momentum for neutrons of $^{12}$C in the JSNS$^2$ simulation, for the
three descriptions of Table~\ref{tab:wigner_facts}. Bands show the $\pm1\sigma$ statistical uncertainty of the MC simulation. The lower panel shows the normalized distribution of the initial nucleon momentum in
the QE events.}
    \label{fig:r_vs_p}
\end{figure}

The second, smaller effect is the intra-shell $r$--$p$ correlation, which is not included in
shell-FACT. 
While the shell-FACT result stays within 2--3\% compared to CORR over most of the $E_m$ range, this is not a trivial result. Indeed, in Fig.~\ref{fig:r_vs_p} we show the mean distance of the interaction point as a function of the initial nucleon momentum for the two neutron shells together. In totR-FACT $\langle r\rangle=2.13$~fm at all momenta. In shell-FACT, $\langle r\rangle$ depends
on $p$ only through the relative contribution of the two shells. It approaches the $1s_{1/2}$ value, 1.65~fm, at low momentum, where the $1p_{3/2}$ momentum distribution vanishes, and rises to 2.30~fm at high momentum. In CORR, $\langle r\rangle$ decreases
from $3.7$~fm at the lowest momenta to a minimum of $1.82$~fm near $300$~MeV. This is akin to the semiclassical picture of a nucleon in a potential at fixed single-particle energy: it moves fast in the interior, where the potential is deep, and slowly near the surface. This is also seen in the kinetic-energy results of Ref.~\cite{Cosyn:2021ber}. 

The main dependence of the radius on the momentum originates from the $1p_{3/2}$ shell, whose
WD is negative in the nuclear interior and positive at the surface
(see Fig.~\ref{fig:C12-proton-Wigner-branches}(d)). It is notable that the $\langle r \rangle$ dependence obtained in the Wigner distribution is significantly different from the shell-FACT approach, they show opposite trends with the mean radius either increasing or decreasing with momentum.
When the momentum is integrated over, as in the comparison to the JSNS$^2$ data of Fig.~\ref{fig:kdar}, the average radii are similar and the results agree within 2--3$\%$. 
The exception is the bin above the $1p_{3/2}$ peak, which in the shell-FACT approach is enhanced by 11\%, while the peak bin is reduced by 2\%. This can be understood from the $r$--$p$ correlations. 
Indeed, note that without rescattering a $1p_{3/2}$ event appears at $E_m=S_{n\alpha}+T_{A-1}$, Eq.~\eqref{eq:Erem}, with
$T_{A-1}\simeq p^2/2M_{A-1}$. The bin that contains the $1p_{3/2}$ peak, i.e. for $16.2<E_m<21.2$~MeV, ends 3.4~MeV above
the separation energy. This bin hence contains the events with $p<263$~MeV, while for $p>263$~MeV they fall in the bin above. 
When the full WD is used the $1p_{3/2}$ nucleons with $p>263$~MeV are found in the interior, at average radius $\langle r\rangle=1.86$~fm, and those with $p<263$~MeV further out,
$\langle r\rangle=2.48$~fm. In the shell-FACT approximation both groups of course have the same average position $\langle r\rangle=2.37$~fm. Since the probability that the proton leaves the nucleus without rescattering decreases with the depth of the interaction point, in shell-FACT, more high-momentum and fewer low-momentum protons leave without rescattering which moves strength from the peak bin to the bin above it.

\begin{table}[t]
\caption{Comparison of the three Wigner descriptions for the JSNS$^2$ missing-energy distribution. The mean radii and momenta are weighted averages over QE events in each shell. }
\label{tab:wigner_kdar_summary}
\begin{ruledtabular}
\begin{tabular}{lcccc}
 & \multicolumn{2}{c}{$\langle r\rangle$ (fm)}
 & \multicolumn{2}{c}{$\langle p\rangle$ (MeV)} \\
Description & $1p_{3/2}$ & $1s_{1/2}$ & $1p_{3/2}$ & $1s_{1/2}$ \\
\hline
CORR            & 2.42 & 1.65 & 180.9 & 143.2 \\
shell-FACT      & 2.37 & 1.65 & 180.9 & 143.2 \\
totR-FACT       & 2.13 & 2.13 & 180.8 & 143.3 \\
\end{tabular}
\end{ruledtabular}
\end{table}

In summary, while the three descriptions differ by less than 2\% in the overall fraction of QE events without rescattering, they differ by up to 11\% in the shape of the $E_m$ distribution: the observable is sensitive to which nucleons rescatter, not only to how many. 
For this comparison, the shell dependence of the interaction point is the dominant effect and the intra-shell $r$--$p$ correlation a secondary one. The former should persist with a more realistic energy smearing and reduced occupations. 
The latter is only visible here because of the sharp relation between $E_m$ and $p$ through the recoil energy, and would be diluted if the removal energy were spread over a few MeV. 

None of the three descriptions reproduces the data in detail\footnote{With the total covariance matrix of
Ref.~\cite{JSNS2:2024bzf}, over the 16 bins used there and with each prediction normalized to unit area, the $\chi^{2}/{\rm ndf}$ values for the three descriptions are: $146.0/16$ (CORR), $143.0/16$ (shell-FACT), and $155.6/16$ (totR-FACT).}. The measured peaks are broader than the predicted ones, as expected for
removal-energies given by Dirac delta functions, and the strength between the peaks is underestimated. The present calculation is by no means a complete description of the JSNS$^2$ measurement, and its purpose is only to quantify the sensitivity to the Wigner initial-state position-momentum phase-space information.

\section{Summary}
\label{sec:summary}

  We have presented the first implementation of a shell-resolved WD description of the
  nuclear initial state in an MC event generator, \textsc{NuWro}. Shell-resolved, angle-averaged proton and neutron WDs for $^{12}$C, $^{16}$O, $^{40}$Ar, and $^{56}$Fe are stored in a tabulated fashion to sample the position and momentum of the struck nucleon together. In contrast to FG models, the Wigner approach provides a joint description of the position and momentum of the bound nucleon while retaining information on the underlying shell structure. The shell-dependent separation energies further allow the binding energy of the struck nucleon to be assigned consistently with the state from which it is removed.

A central difficulty in using finite-nucleus WDs in an MC generator is that they are not positive definite. We developed two complementary implementations to address this problem. In the weighted-event approach, the original signed WD is retained by sampling separately from its positive and negative components and assigning the corresponding sign and normalization correction to the event weight. This preserves the complete WD information, at the cost of a reduced statistical efficiency.  For \textsc{NuWro}'s standard unweighted event generation, we instead construct a non-negative coarse-grained distribution. The procedure absorbs negative regions into neighboring positive strength and restores the shell momentum distribution exactly. The resulting distribution can be sampled without introducing signed event weights.

The consequences of the Wigner initial state were studied using electron and neutrino scattering data. At the inclusive level, the WD cross section reduces to the PWIA. At the exclusive level, the position--momentum correlation affects observables through the INC. We isolated it in the missing-energy distribution measured by JSNS$^2$, by comparing the original WD with two factorized constructions that give identical primary vertex distributions. The $E_m$ distribution is mainly sensitive to the shell dependence of the interaction point that the WD provides: removing it changes the heights of the shell peaks by
7--10\%. The intra-shell $r$--$p$ correlation, which places fast nucleons in the interior and slow nucleons further out, is a smaller effect, confined to the $1p_{3/2}$ peak region. This shell dependence does not rely on the delta-function removal energies of the
present model and should persist with realistic spectral functions. It should be taken into account when these data are compared with models in which the interaction
point does not depend on the shell.

Several extensions follow naturally. Spectroscopic factors and broadened
removal-energy distributions can be included without changing the radial sampling, which also
provides position information for calculations based on hole SFs. The WD offers a phase-space description of the occupied
states that can be directly used for a consistent Pauli blocking recipe of the outgoing nucleon. The WD information can also be exported into the distributions of secondary target nucleons in the INC. These developments will be discussed in future work.

\begin{acknowledgments}
We thank Eric Marzec for details of the JSNS$^2$ experiment signal definition.  This work is partly (K.M.G., B.E.K., and J.T.S.) or fully (R.D.B., J.L.B., and H.P.) supported by the National Science Center, Poland under grant UMO-2021/41/B/ST2/02778.
A.N. is supported by the Neutrino Theory Network (NTN) under Award Number DE-AC02-07CH11359.
\end{acknowledgments}

\appendix

\makeatletter
\@addtoreset{figure}{section}
\@addtoreset{table}{section}
\makeatother

\renewcommand{\thefigure}{\Alph{section}\arabic{figure}}
\renewcommand{\thetable}{\Alph{section}\arabic{table}}

\section{Shell separation energies}
\label{app:separation}

The shell-dependent separation energies $S_{\tau\alpha}$ and occupation numbers \(N_{\tau\alpha}\) used in the WD implementation are collected in Table~\ref{tab:separation_energies}.

\begin{table}[ht]
\caption{Shell-dependent occupation numbers $N_{\tau\alpha}$ and separation energies $S_{\tau\alpha}$ in MeV. 
}
\label{tab:separation_energies}
\begin{ruledtabular}
\begin{tabular}{cccccc}
Nucleus & $\alpha$ & $N_{p\alpha}$ & $N_{n\alpha}$
        & $S_{p\alpha}$ & $S_{n\alpha}$ \\

\hline

\multirow{2}{*}{$^{12}\mathrm{C}$}
& $1s_{1/2}$ & 2 & 2 & 40.5 & 44.4 \\
& $1p_{3/2}$ & 4 & 4 & 14.5 & 17.8 \\
\hline
\multirow{3}{*}{$^{16}\mathrm{O}$}
& $1s_{1/2}$ & 2 & 2 & 36.2 & 40.3 \\
& $1p_{3/2}$ & 4 & 4 & 16.7 & 20.6 \\
& $1p_{1/2}$ & 2 & 2 & 8.8  & 12.5 \\
\hline
\multirow{7}{*}{$^{40}\mathrm{Ar}$}
& $1s_{1/2}$ & 2 & 2 & 43.5 & 47.7 \\
& $1p_{3/2}$ & 4 & 4 & 30.5 & 33.8 \\
& $1p_{1/2}$ & 2 & 2 & 26.5 & 29.8 \\
& $1d_{5/2}$ & 6 & 6 & 17.7 & 20.2 \\
& $2s_{1/2}$ & 2 & 2 & 12.8 & 15.7 \\
& $1d_{3/2}$ & 2 & 4 & 11.0 & 13.3 \\
& $1f_{7/2}$ & -- & 2 & -- & 7.5 \\
\hline
\multirow{9}{*}{$^{56}\mathrm{Fe}$}
& $1s_{1/2}$ & 2 & 2 & 45.4 & 53.3 \\
& $1p_{3/2}$ & 4 & 4 & 33.3 & 40.3 \\
& $1p_{1/2}$ & 2 & 2 & 30.6 & 37.5 \\
& $1d_{5/2}$ & 6 & 6 & 21.0 & 27.3 \\
& $2s_{1/2}$ & 2 & 2 & 16.1 & 22.2 \\
& $1d_{3/2}$ & 4 & 4 & 15.5 & 21.4 \\
& $1f_{7/2}$ & 6 & 8 & 9.1  & 14.7 \\
& $2p_{3/2}$ & --    & 2 & --   & 9.8 \\

\end{tabular}
\end{ruledtabular}
\end{table}

\begin{figure*}[t]
    \centering
\includegraphics[width=\textwidth]{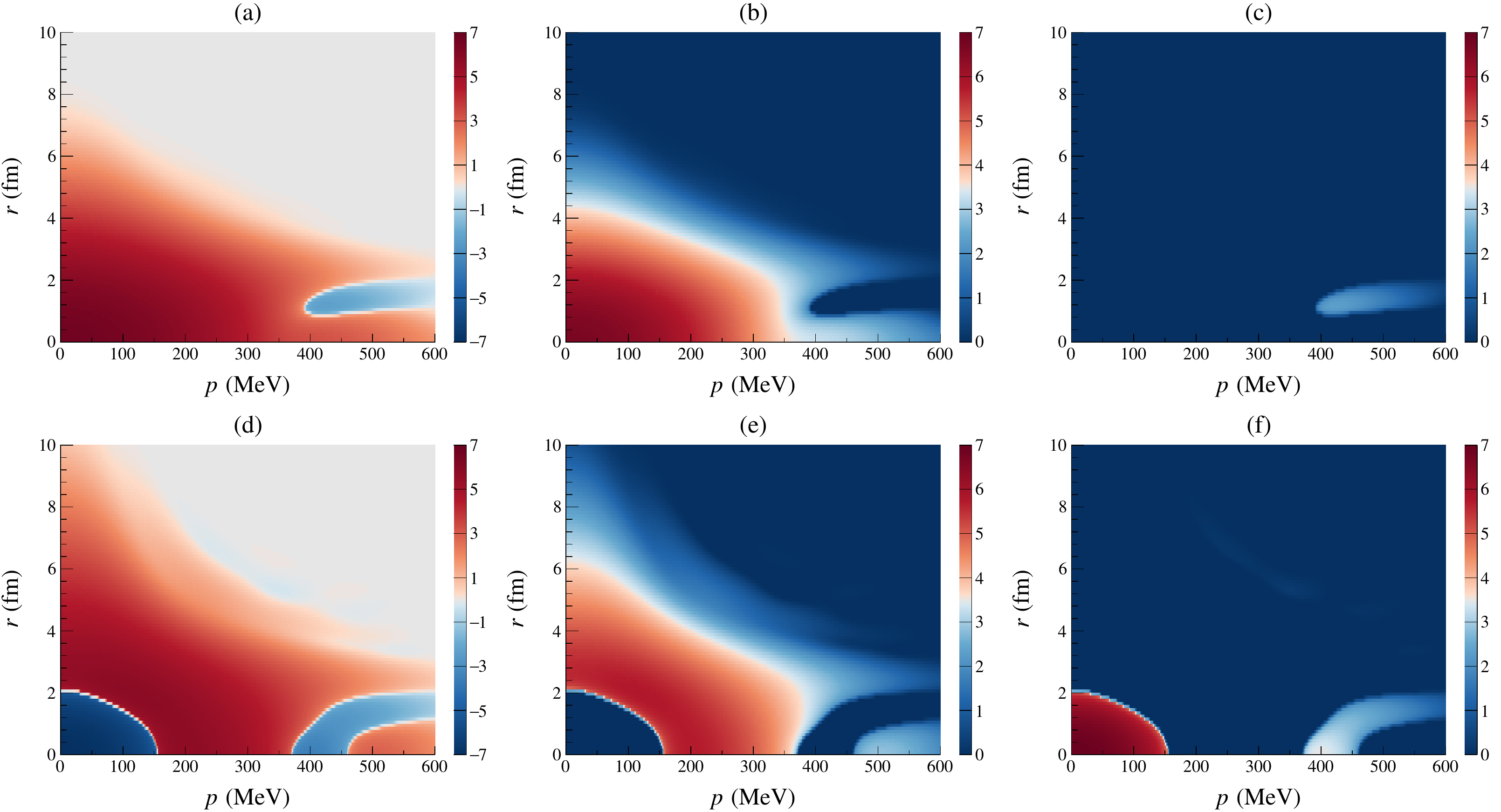}
    \caption{
        Proton shell-resolved WDs for \(^{12}\mathrm{C}\).
        Panels (a)--(c) correspond to \(1s_{1/2}\) shell, while
        panels (d)--(f) correspond to \(1p_{3/2}\) shell.
        Panels (a) and (d) show the original signed distributions
        \(w_{\tau\alpha}(r,p)\).
        Panels (b) and (e) show the positive components
        \(w_{\tau\alpha}^{+}(r,p)\), while panels (c) and
        (f) show magnitudes of the negative components
        \(w_{\tau\alpha}^{-}(r,p)\). The color scale is logarithmic:
$z=\mathrm{sgn}(w)\log_{10}(1+|w|/w_0)$ in panels (a) and (d), and
$z_\pm=\log_{10}(1+w^\pm/w_0)$ in the other panels, with $w_0=10^{-7}w_{\max}$
and $w_{\max}$ the largest $|w_{\tau\alpha}|$ of the shown shells, so that
$|z|=7$ corresponds to $w_{\max}$. Thus, small values are strongly enhanced.
    }
    \label{fig:C12-proton-Wigner-branches}
\end{figure*}

\section{Monte Carlo implementation}
\label{app:wigner_implementation}

\subsection{Weighted mode}
\label{app:weighted_wigner}

In the weighted event mode, we retain the original information from the signed WD which is negative at some regions in the phase-space.

\subsubsection{Branching}

For each shell, the distribution is separated into its positive and negative branches,

\begin{align}
w_{\tau\alpha}^{+}(r,p)
&=\frac{1}{2}\left[w_{\tau\alpha}(r,p)+\left|w_{\tau\alpha}(r,p)\right|\right],
\label{eq:def_pos}
\\
w_{\tau\alpha}^{-}(r,p)
&=\frac{1}{2}\left[\left|w_{\tau\alpha}(r,p)\right|-w_{\tau\alpha}(r,p)\right],
\label{eq:def_neg}
\end{align}
such that
\begin{equation}
w_{\tau\alpha}(r,p)
=
w_{\tau\alpha}^{+}(r,p)-w_{\tau\alpha}^{-}(r,p).
\label{eq:posneg}
\end{equation}

Both $w_{\tau\alpha}^{+}$ and $w_{\tau\alpha}^{-}$ are non-negative and can be used to construct MC sampling distributions. The two branches of the $^{12}$C proton shells are shown in Fig.~\ref{fig:C12-proton-Wigner-branches}.

The integrated strengths of the two branches are defined as
\begin{equation}
N^{\pm}_{\tau\alpha}=\int d^3r\int\frac{d^3p}{(2\pi)^3}\,w^{\pm}_{\tau\alpha}(r,p),
\end{equation}

The total positive and negative strengths are

\begin{equation}
    N_{\tau}^{\pm} = \sum_{\alpha\in\mathcal{S}_\tau}
N_{\tau\alpha}^{\pm}.
\end{equation}
The negative fraction of Eq.~\eqref{eq:neg_fraction} is $f_-^\tau=N_\tau^-/(N_\tau^++N_\tau^-)$, listed in Table~\ref{tab:wigner_negative_fraction}.

\subsubsection{Sampling}

Once the isospin of the struck nucleon is known from the nature of the interaction, the shell and the branch are sampled jointly according to

\begin{equation}
P(\alpha,\pm\mid\tau)
=
\frac{
N_{\tau\alpha}^{\pm}
}{
 N_{\tau}^{\rm ABS}
}.
\label{eq:wigner_shell_branch_sampling}
\end{equation}

where,

\begin{equation}
 N_{\tau}^{\rm ABS} = N_{\tau}^{+}+N_{\tau}^{-} 
 \label{eq:N_abs}
\end{equation}

The sign associated with the sampled branch is defined as

\begin{equation}
s =
\begin{cases}
+1, & \text{for the positive branch},\\
-1, & \text{for the negative branch}.
\end{cases}
\end{equation}

After selecting $(\alpha,s)$, the radial position is sampled from
\begin{equation}
P^{\pm}_{\tau\alpha}(r)=\frac{4\pi r^2}{N^{\pm}_{\tau\alpha}}\int\frac{d^3p}{(2\pi)^3}\,w^{\pm}_{\tau\alpha}(r,p).
\end{equation}

The momentum magnitude is subsequently sampled from the conditional distribution
\begin{equation}
P_{\tau\alpha}^{\pm}(p\mid r)
=
\frac{
4\pi p^2w_{\tau\alpha}^{\pm}(r,p)
}{
\displaystyle\int d^3p'\,w_{\tau\alpha}^{\pm}(r,p')
}.
\end{equation}
After sampling the sign, shell, position, and momentum, the binding energy is assigned according to Eq.~\eqref{eq:B_WD}.

\subsubsection{Wigner sign and normalization}

The phase-space point is sampled from the non-negative branches of Eqs.~\eqref{eq:def_pos} and ~\eqref{eq:def_neg}, whose combined normalization is $N^{\rm ABS}_\tau$. The original signed WD, however, is normalized by

\begin{equation}
N_{\tau}^{\mathrm{SIGN}}
=
N_{\tau}^{+}
-
N_{\tau}^{-} \equiv  N_\tau.
\label{eq:wigner_signed_strength}
\end{equation}

The difference between the absolute sampling normalization and the signed physical normalization is corrected by

\begin{equation}
C_{\tau}
=
\frac{
N_{\tau}^{\mathrm{ABS}}
}{
N_{\tau}^{\mathrm{SIGN}}
}.
\label{eq:wigner_normalization_factor}
\end{equation}

With Eqs.~\eqref{eq:N_abs} and \eqref{eq:wigner_signed_strength}, $C_\tau=1/(1-2f_-^\tau)$, so $C_\tau$ follows from Table~\ref{tab:wigner_negative_fraction}. If $\omega_j^{(0)}$ denotes the unsigned event  weight calculated for the sampled kinematics, the Wigner event weight is defined as

\begin{equation}
\omega_j
=
s_j
C_{\tau_j}
\omega_j^{(0)},
\,
s_j
=
\pm 1.
\label{eq:wigner_signed_event_weight}
\end{equation}

The physical prediction is obtained only after taking into account the two contributions,

\begin{equation}
\sigma
=
\sigma^{(+)}-\sigma^{(-)},
\label{eq:cross_section_decomposition}
\end{equation}

where \(\sigma^{(+)}\) and \(\sigma^{(-)}\) denote the contributions obtained from the positive and negative branches, respectively, and \(\sigma\) is the total physical cross section.

\subsubsection{Differential cross sections and statistical precision}

\begin{figure}[!t]
    \centering
\includegraphics[width=\linewidth]{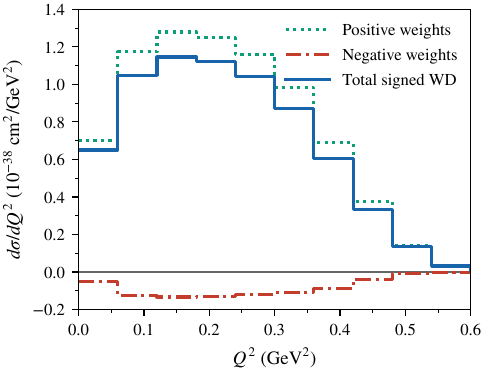}
    \caption{Differential cross section in \(Q^2\) for 500 MeV \(\nu_\mu\) QE scattering on $^{40}$Ar. The overall result is decomposed according to Eq.~\eqref{eq:cross_section_decomposition}.
}
    \label{fig:wigner_q2_weight_decomposition}
\end{figure}

In the weighted mode, differential cross sections are evaluated as
\begin{equation}
  \left(\frac{d\sigma}{dX}\right)_k = \frac{1}{N_{\rm gen}\,\Delta X_k}
  \sum_{j=1}^{N_{\rm gen}} \omega_j\,\chi_k(X_j),
\end{equation}

where $k$ labels the bins of the variable $X$, $\Delta X_k$ is the bin width, $\chi_k(X_j)=1$ if the value $X_j$ of event $j$ falls in bin $k$ and 0 otherwise: summed over all bins this reproduces Eq.~\eqref{eq:total_sigma}. For every variable fixed at the primary vertex, the general argument of Sec.~\ref{subsec:lfg_cross_section} applies: in a limit of large statistics (\textit{i.e.} small fluctuations) all the differential cross sections must be positive. An illustration of this can be seen in Fig.~\ref{fig:wigner_q2_weight_decomposition}.

Each event weight carries the sign $s_j$ of Eq.~\eqref{eq:wigner_signed_event_weight}. The signed weights partly cancel each other but their statistical fluctuations do not. Cross sections evaluated with weighted-mode samples has statistical precision of a smaller, effective sample. Its size,

\begin{equation}
  N_{\rm eff}=\frac{\left(\sum_j\omega_j\right)^2}{\sum_j\omega_j^2},
\end{equation}

is such that the relative statistical uncertainty of a bin is $1/\sqrt{N_{\rm eff}}$,
with the sums running over the events in that bin. The signs leave
$\sum_j\omega_j^2$ unchanged and reduce only $\sum_j\omega_j$. The
negative-branch events carry the fraction $f_-^\tau$ of $\sum_j|\omega_j|$, so that $\sum_j\omega_j$ is
reduced by the factor $1-2f_-^\tau=1/C_\tau$. For the full sample, $N_{\rm eff}$ is then
$C_\tau^2$ times smaller than for the same events without signs, and a given precision requires $C_\tau^2$ times as many events: 17\% more for $^{12}$C, 20\% for $^{16}$O, 47\% and 51\% for $^{40}$Ar protons and neutrons, and 56\% and 62\% for $^{56}$Fe protons and neutrons. The same quantity
fixes the minimal statistics needed to avoid negative bin contents: a bin can become negative
only through a downward fluctuation of $\sqrt{N_{\rm eff}}$ standard deviations, so bins with $N_{\rm eff}\gtrsim9$ stay positive at the $3\sigma$ level.

\subsection{Unweighted mode}
\label{app:cg_wigner}

The weighted implementation provides a direct way to retain the full signed WD. However, the most common event-generation mode in \textsc{NuWro} is based on unweighted events generated with the acceptance-rejection algorithm. Since the original WD is not positive definite, it cannot be sampled directly in this mode. Thus, we construct a non-negative coarse-grained distribution that can be used for unweighted event generation while preserving the relevant properties of the original distribution.

\subsubsection{Coarse-graining algorithm}
\label{subsubsec:coarse_graining}

To use the WD in the unweighted mode one needs a non-negative distribution. Hence, we construct a coarse-grained distribution \(
w_{\tau\alpha}^{\mathrm{CG}}(r,p)\geq 0,
\) by absorbing the negative strength of the original signed WD into the
adjacent positive strength, along one phase-space direction at a time. Consider a one-dimensional slice of the WD and define
\begin{equation}
  Y_{\tau\alpha}(x;\xi)=x^2\,w_{\tau\alpha}(r,p),
  \label{eq:Ydef}
\end{equation}

where \(x\) denotes the coordinate along which the coarse-graining is
performed and $\xi$ the coordinate held fixed, i.e.
\begin{equation}
  (x;\xi) \;=\;
  \begin{cases}
    (r;p), & \text{radial coarse-graining},\\[2pt]
    (p;r), & \text{momentum coarse-graining}.
  \end{cases}
  \label{eq:cgchoice}
\end{equation}

The $x^{2}$ weight is used throughout because it is $x^{2}w_{\tau\alpha}$, and
not $w_{\tau\alpha}$ itself, that carries the phase-space measure of the distribution from which events are sampled.  Imposing non-negativity on the weighted variable guarantees a valid sampling density. Intervals of consecutive grid points with $Y_{\tau\alpha}(x;\xi)<0$ are first
identified and treated in order of increasing $x$. Each interval is the seed of
a block $B$, which is enlarged one grid point at a time. At each step the
unused grid points immediately to the left and to the right of the current
block are considered, and the one with the larger non-negative value of $Y_{\tau\alpha}$ is added. The enlargement stops as soon as
the signed sum over the block becomes non-negative,
\begin{equation}
  I_{B,\tau\alpha}(\xi)=\sum_{x_i\in B}Y_{\tau\alpha}(x_i;\xi) \ge 0,
\end{equation}
or until no admissible neighbor is left. For a block \(B\), we define its original positive strength as
\begin{equation}
  I^{+}_{B,\tau\alpha}(\xi)=\sum_{x_i\in B}\max\!\left[Y_{\tau\alpha}(x_i;\xi),0\right],
\end{equation}

and retain only that positive part, rescaled by a common factor,

\begin{equation}
Y^{\mathrm{CG}}_{B,\tau\alpha}(x;\xi)=
\frac{\max\!\left[I_{B,\tau\alpha}(\xi),0\right]}{I^{+}_{B,\tau\alpha}(\xi)}\,
\max\!\left[Y_{\tau\alpha}(x;\xi),0\right],
  \label{eq:blockcg}
\end{equation}

so that

\begin{equation}
  Y^{\rm CG}_{B,\tau\alpha}\ge0,\quad
  \sum_{x_i\in B}Y^{\rm CG}_{B,\tau\alpha}(x_i;\xi)=\max\!\left[I_{B,\tau\alpha}(\xi),0\right].
\end{equation}

The negative strength of the block is absorbed by a local reduction of its positive strength, at fixed block integral. Bins that are not utilized in the coarse-graining algorithm are retained unchanged. Assembling all blocks gives the reconstructed
one-dimensional slice
\begin{equation}
  Y^{\mathrm{CG}}_{\tau\alpha}(x;\xi) \;=\;
  Y^{\mathrm{CG}}_{B,\tau\alpha}(x;\xi) ,
  \,x\in B ,
  \label{eq:sliceassembly}
\end{equation}

where the set of blocks $B$, including single grid points, covers the complete slice. A block can run into the end of the grid or into the next negative interval or its block before that. The residual negative strength of such a
block is dropped. In total, only about $0.1\%$ of the negative strength is discarded in this way. Since the discarded strength is not exactly zero, the reconstructed slice does not automatically carry the signed normalization of the original slice.
Defining

\begin{equation}
  I^{\rm ORIG}_{\tau\alpha}(\xi)=\sum_i Y_{\tau\alpha}(x_i;\xi)\,,\,
  I^{\rm CG}_{\tau\alpha}(\xi)=\sum_i Y^{\rm CG}_{\tau\alpha}(x_i;\xi),
\end{equation}

the complete slice is rescaled by
\begin{equation}
  \lambda_{\tau\alpha}(\xi) \;=\;
  \frac{I^{\mathrm{ORIG}}_{\tau\alpha}(\xi)}{I^{\mathrm{CG}}_{\tau\alpha}(\xi)} ,
 \
  \widetilde{Y}^{\mathrm{CG}}_{\tau\alpha}(x;\xi) \;=\;
  \lambda_{\tau\alpha}(\xi)\, Y^{\mathrm{CG}}_{\tau\alpha}(x;\xi) ,
  \label{eq:lambda}
\end{equation}

which restores the original signed slice normalization exactly. The quantity
$1-\lambda_{\tau\alpha}$
is the discarded negative strength relative to the positive strength kept in
the slice. Averaged over all shell slices of the four nuclei without weighting,
$\langle 1-\lambda\rangle=0.012$. Weighted by the shell radial density, the
average is below $10^{-3}$, because $\lambda$ differs appreciably from unity
almost only where the shell's own density vanishes. The corresponding coarse-grained distribution is recovered according to

\begin{equation}
\widetilde{w}_{\tau\alpha}^{\mathrm{CG}}(r,p)
=
\frac{\widetilde{Y}_{\tau\alpha}^{\mathrm{CG}}(x;\xi)}{x^2}.
\label{eq:cg_back_transform}
\end{equation}

The coarse-graining procedure cannot, in general, preserve both the
radial and momentum marginals of the original signed WD while simultaneously requiring the reconstructed distribution to remain non-negative. A choice must be made as to which marginal should be retained exactly. We choose to preserve the momentum distribution because the initial-nucleon momentum enters directly into the primary-interaction kinematics and the elementary neutrino--nucleon cross section. For each shell
$\alpha$, the coarse-graining procedure is applied to the
momentum-dependent profile at every fixed value of $r$, as in
Eq.~(\ref{eq:Ydef})
\begin{equation}
Y_{\tau\alpha}(p;r)=p^2\,w_{\tau\alpha}(r,p).
\label{eq:cg_momentum_profile}
\end{equation}

This produces an intermediate non-negative distribution
\(\widetilde{w}_{\tau\alpha}^{\mathrm{CG}}(r,p)\). At every fixed value of \(r\), the block reconstruction preserves the integrated strength of the original profile,
\begin{equation}
\int d^3p\,
\widetilde{w}_{\tau\alpha}^{\mathrm{CG}}(r,p)
=
\int d^3p\, w_{\tau\alpha}(r,p).
\label{eq:cg_slice_normalization}
\end{equation}

After all slices have been combined, the momentum marginal of the intermediate table does not necessarily reproduce that of the original WD. The corresponding intermediate coarse-grained momentum marginal is
\begin{equation}
\widetilde{n}_{\tau\alpha}^{\mathrm{CG}}(p)
=
\int d^3r\,
\widetilde{w}^{\rm CG}_{\tau\alpha}(r,p).
\end{equation}

The intermediate table is corrected using the non-negative factor
\begin{equation}
\eta_{\tau\alpha}(p)
=
\frac{
n_{\tau\alpha}(p)
}{
\widetilde{n}_{\tau\alpha}^{\mathrm{CG}}(p)
},
\label{eq:cg_momentum_scale}
\end{equation}
giving the final coarse-grained distribution
\begin{equation}
w_{\tau\alpha}^{\mathrm{CG}}(r,p)
=
\eta_{\tau\alpha}(p)\,
\widetilde{w}_{\tau\alpha}^{\mathrm{CG}}(r,p).
\label{eq:cg_final}
\end{equation}

\subsubsection{Sampling}

The resulting distribution is non-negative by construction and reproduces the original shell momentum distribution exactly. Event generation is performed separately for protons and neutrons. The shell is first selected according to its strength,
\begin{equation}
P(\alpha\mid\tau)
=
\frac{N_{\tau\alpha}^{\mathrm{CG}}}
{\displaystyle\sum_{\beta\in\mathcal{S}_\tau}N_{\tau\beta}^{\mathrm{CG}}},
\,
N^{\rm CG}_{\tau\alpha}=\int d^3r\int\frac{d^3p}{(2\pi)^3}\,w^{\rm CG}_{\tau\alpha}(r,p).
\label{eq:cg_shell_sampling}
\end{equation}

Once the shell has been selected, the interaction point is sampled from the radial marginal,
\begin{equation}
P^{\rm CG}_{\tau\alpha}(r)=\frac{4\pi r^2}{N^{\rm CG}_{\tau\alpha}}\int\frac{d^3p}{(2\pi)^3}\,w^{\rm CG}_{\tau\alpha}(r,p),
\label{eq:cg_radius_sampling}
\end{equation}
The momentum magnitude is sampled at the selected radius from
\begin{equation}
P^{\rm CG}_{\tau\alpha}(p\,|\,r)=\frac{4\pi p^2\,w^{\rm CG}_{\tau\alpha}(r,p)}{\int d^3p'\,w^{\rm CG}_{\tau\alpha}(r,p')}.
\label{eq:cg_momentum_conditional}
\end{equation}

\begin{figure}[t]
    \centering
\includegraphics[width=\columnwidth]{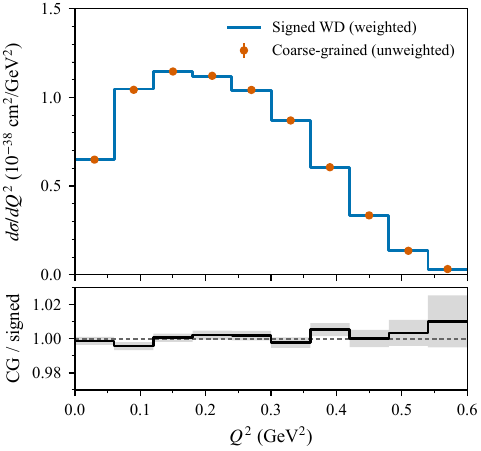}
    \caption{Differential cross section in $Q^{2}$ for  $500~\mathrm{MeV}$ $\nu_{\mu}$ scattering on $^{40}\mathrm{Ar}$. The signed WD (solid blue line) refers to weighted-event mode, while the coarse-grained distribution (orange points) corresponds to  unweighted-event mode. The lower panel shows the ratio of the coarse-grained distribution to the signed WD result, with its MC \(\pm1\sigma\) statistical uncertainty band.
    }
\label{fig:wigner_q2_weighted_unweighted}
\end{figure}

The binding energy is then assigned as in Eq.~\eqref{eq:B_WD}. Because the momentum marginal is preserved, the integrated strength of each coarse-grained shell is equal to the strength of the corresponding original WD,

\begin{equation}
N^{\rm CG}_{\tau\alpha}=\int\frac{d^3p}{(2\pi)^3}\,n_{\tau\alpha}(p)=N^{+}_{\tau\alpha}-N^{-}_{\tau\alpha}=N^{\rm SIGN}_{\tau\alpha},
\label{eq:cg_shell_normalization}
\end{equation}

where

\begin{equation}
 \sum_{\alpha \in \mathcal{S}_\tau} N^{\rm SIGN}_{\tau\alpha} = N^{\rm SIGN}_{\tau} 
\end{equation}

Hence, for the unweighted mode neither \(s\) nor \(C_\tau\) is required.

\subsection{Performance}
\label{app:performance}

By construction, the coarse-graining introduces distortion to the radial density. Hence, the coarse-grained construction should be regarded as an approximate representation of the original signed WD rather than as an
equivalent transformation. To quantify the radial distortion, we first define the radial probability density as

\begin{equation}
P_{\tau}(r)
=
\frac{4\pi r^{2}}{N_\tau}\rho_{\tau}(r).
\label{eq:cg_radial_probability}
\end{equation}

The induced distortion is then quantified using the normalized \(L_1\) distance defined as
\begin{equation}
d_{r}^{\tau}
=
\frac{
\displaystyle
\int dr\,
\left|
P_{\tau}^{\mathrm{CG}}(r)
-
P_{\tau}^{\mathrm{ORIG}}(r)
\right|
}{
\displaystyle
\int dr\,
\left|
P_{\tau}^{\mathrm{ORIG}}(r)
\right|
},
\label{eq:cg_radial_L1}
\end{equation}

\begin{table}[t]
\caption{
Normalized $L^{1}$ distances of \(P_\tau(r)\) induced by the shell-resolved coarse-grained construction.
}
\label{tab:momentum_cg_radial_distortion}
\begin{ruledtabular}
\begin{tabular}{lcc}
Target
&
$d_{r}^{p}$ [\%]
&
$d_{r}^{n}$ [\%]
\\
\hline
$^{12}\mathrm{C}$  & 2.97 & 2.90 \\
$^{16}\mathrm{O}$  & 3.08 & 3.02 \\
$^{40}\mathrm{Ar}$ & 6.88 & 6.34 \\
$^{56}\mathrm{Fe}$ & 5.92 & 7.24 \\
\end{tabular}
\end{ruledtabular}
\end{table}

where $P_{\tau}^{\mathrm{ORIG}}(r)$ is obtained by summing the signed shell WDs, and $P_{\tau}^{\mathrm{CG}}(r)$ is
obtained from the corresponding sum of the coarse-grained shell distributions. As shown in Table~\ref{tab:momentum_cg_radial_distortion}, the induced distortion of the radial marginal is $\sim3\%$ for $^{12}$C and $^{16}$O and increases to $\sim7\%$ for $^{40}$Ar and $^{56}$Fe. This quantifies the modification of the position information required to replace the signed WD by a non-negative distribution while preserving the marginal momentum distribution. We further test whether the coarse-grained construction reproduces the prediction obtained by the original signed WD at the cross-section level. We compare, for $^{40}$Ar, the differential cross-section in four momentum transfer squared, $Q^2$, obtained with the two modes. As shown in Fig.~\ref{fig:wigner_q2_weighted_unweighted}, the weighted-event mode calculation based on the original signed WD and the unweighted-event mode calculation based on the coarse-grained distribution give nearly identical $Q^{2}$ distributions within the level of statistical fluctuation.

\bibliography{main}

\end{document}